\documentclass[a4paper,12pt]{article}
\usepackage[pctex32]{graphics}
\usepackage{amssymb,amsmath}
\usepackage{amsmath,amssymb}
\usepackage{latexsym}
\usepackage{epsfig}
\usepackage[english]{babel}

\newcommand{\be}{\begin{equation}}
\newcommand{\ee}{\end{equation}}
\newcommand{\ba}{\begin{eqnarray}}
\newcommand{\ea}{\end{eqnarray}}

\begin{document}

\begin{titlepage}

\vspace{5mm}

\begin{center}

{\Large \bf Stability issues of  black hole  in non-local  gravity}

\vskip .6cm

{Yun Soo Myung$^a$\footnote{e-mail address: ysmyung@inje.ac.kr} and Young-Jai Park$^b$\footnote{e-mail address:yjpark@sogang.ac.kr}}\\[8mm]

{$^a$Institute of Basic Sciences and Department  of Computer Simulation, \\Inje University, Gimhae 50834, Korea\\[0pt]}
{$^b$Department of Physics, Sogang University, Seoul 04107, Korea\\[0pt]}
\vskip .6cm

\end{center}

\begin{abstract}

We discuss  stability issues of Schwarzschild black hole in non-local gravity.
It is shown that the stability analysis of black hole  for the unitary and renormalizable non-local gravity with  $\gamma_2=-2\gamma_0$ cannot be performed in the Lichnerowicz operator approach.
On the other hand, for  the unitary and non-renormalizable case with  $\gamma_2=0$, the black hole is stable against the metric perturbations.
For non-unitary and renormalizable  local gravity with  $\gamma_2=-2\gamma_0={\rm const}$ (fourth-order gravity),  the small  black holes are  unstable against the metric perturbations.
This implies that what makes the problem difficult in stability analysis of black hole is the simultaneous requirement of  unitarity and renormalizability around the Minkowski spacetime.

\end{abstract}

 \vskip .6cm

\vskip 0.8cm

\vspace{15pt} \baselineskip=18pt

\thispagestyle{empty}
\end{titlepage}

\newpage
\section{Introduction}

It turns out that the infinite derivative gravity (non-local gravity) is ghost-free and renormalizable around the Minkowski spacetime background when one chooses the exponential form of an  entire function
~\cite{Modesto:2011kw,Biswas:2011ar}. We note that  renormalizability can be easily checked by showing  the finiteness of the Newtonian potential at the origin
from the propagator~\cite{Accioly:2013hwa,Modesto:2014eta,Giacchini:2016xns,Accioly:2017xmm}.

On the other hand, all Ricci-flat spacetimes including Schwarzschild black hole  are exact solutions for  non-local gravitational theories~\cite{Li:2015bqa}.
In order to check  that  the Schwarzschild black hole exists really in  the unitary (ghost-free) and renormalizable  non-local gravity, one has to perform the stability analysis of the black hole.
If the black hole solution  passes the stability test, one may  save that black hole.
Recently, it has shown that  the Schwarzschild black hole is stable against  linear perturbations  for a subclass of unitary  non-local gravity with $\gamma_2=0$~\cite{Calcagni:2017sov}.
However, this case is not  a renormalizable gravity around Minkowski spacetime.
Although  the  non-locality (operators with infinitely many derivatives) is needed to have a ghost-free and renormalizable gravity,
the presence of higher derivative gravity may make the black hole unsustainable.  This implies that non-locality may not be a good tool to cure the black hole solutions.

In this work, we  wish to discuss  stability issues of Schwarzschild black hole in non-local gravity.
We derive a linearized equation (\ref{EOM4}) which governs the stability of black hole for the unitary and renormalizable non-local gravity with  $\gamma_2=-2\gamma_0$.
However, we could not perform the stability analysis of black hole in the Lichnerowicz operator approach because the appearance  of entire function in the linearized equation (\ref{EOM11}).
On the other hand, for a unitary and non-renormalizable gravity with  $\gamma_2=0$, it has shown that  the black hole is stable against the metric perturbations.
This is possible  because  this case reduces to the Einstein gravity or $f(R)$ gravity~\cite{Starobinsky:1980te}, which are surely independent of the entire function~\cite{Calcagni:2017sov}.
Next, for the non-unitary and renormalizable local   gravity with  $\gamma_2=-2\gamma_0={\rm const}$ (fourth-order gravity)~\cite{Stelle:1976gc}, using the Gregory-Laflamme black string instability~\cite{Gregory:1993vy},
the small  black holes are  unstable against the Ricci tensor  perturbations. This contrasts to the conventional  stability analysis of black hole  in Einstein gravity or $f(R)$ gravity.
It implies  that the simultaneous requirement for unitarity and renormalizability  makes the stability analysis  difficult.

\section{Non-local gravity}
 A non-local   gravity in four dimensions is generally defined by~\cite{Calcagni:2017sov}
 \begin{equation} \label{action1}
S_g = \frac{2}{\kappa^2}\int  d^4 x \sqrt{|g|} \left[R + R \gamma_0(\square) R+R_{\mu\nu} \gamma_2(\square) R^{\mu\nu} + V_g\right],
 \end{equation}
where $\kappa^{2}=32\pi G$, the d'Alembertian $\square=g^{\mu\nu}\nabla_{\mu}\nabla_{\nu}$,  and the potential term $V_g$ is at least cubic in the curvature and at least quadratic in the Ricci tensor.
Hereafter, we choose $V_g=0$ for simplicity.
The non-local gravity with $\gamma_2=0$ is unitary and non-renormalizable around Minkowski spacetime when choosing  one of the two  form factors
\begin{eqnarray}
\gamma_0 &\hspace{-0.2cm}=\hspace{-0.2cm}& - \frac{e^{H(\square)} - 1}{6\square}\,, \label{ff1}\\
\gamma_0 &\hspace{-0.2cm}=\hspace{-0.2cm}& - \frac{e^{H(\square)} \left(1-\frac{\square}{M^2} \right) - 1}{6 \square}\,, \label{ff2}
\end{eqnarray}
where $H(\Box)$ is an entire function and $M$ is a mass scale.
 The first form factor was first proposed by Biswas, Mazumdar and Siegel  with $H(\square)=\square$ \cite{Biswas:2005qr},
 whereas the second one appears in the non-local extension of Starobinsky gravity \cite{Briscese:2013lna}.

 For $\gamma_2=-2\gamma_0$, the tree-level unitarity analysis  for the metric perturbation around Minkowski spacetime shows that  the graviton propagator for (\ref{ff1})  takes the form of
 $\Pi(k)=e^{-H(-k^2)}\Pi_{\rm GR}$~\cite{Modesto:2014eta}. Furthermore, its renormalizability can be easily seen by computing the Newtonian potential from this propagator.

Before we proceed, we would like to note that  the unitary  case of $\gamma_2 =0$  was reduced to  the stability analysis of the Schwarzschild black hole in Einstein gravity for the case of (\ref{ff1}) and  $f(R)$ gravity for
 the case of (\ref{ff2})~\cite{Calcagni:2017sov}.
However, this case is not  a renormalizable gravity when quantizing  around Minkowski spacetime. Therefore, in order to make a connection to the unitary and renormalizable non-local gravity, one considers the case of  $\gamma_2=-2\gamma_0$ in the beginning of stability analysis for the black hole.

\section{Equation of motion: Ricci-flat solutions}
 The equation of motion  is derived from the action (\ref{action1}) as~\cite{Mirzabekian:1995ck}
\begin{eqnarray}
E_{\mu\nu} &\equiv& G_{\mu\nu} - \frac{1}{2} g_{\mu\nu} R \gamma_0(\square) R -  \frac{1}{2} g_{\mu\nu} R_{\alpha \beta} \gamma_2(\square) R^{\alpha \beta}
+2\frac{\delta R}{\delta g^{\mu \nu}} \gamma_0(\square)R \nonumber \\
&+& \frac{\delta R_{\alpha \beta}}{\delta g^{\mu \nu}} \gamma_2(\square)R^{\alpha \beta}
 + \frac{\delta R^{\alpha \beta}}{\delta g^{\mu \nu}  } \gamma_2(\square)R_{\alpha \beta} +  \frac{\delta \square^r}{\delta g^{\mu\nu} }
 \left[\frac{\gamma_0(\square^l)-\gamma_0(\Box^r)}{\square^r - \square^l} R^2 \right]\nonumber \\
&+& \frac{\delta \square^r}{\delta g^{\mu\nu}}\left[
  \frac{ \gamma_2(\square^l)- \gamma_2(\square^r)}{\square^r - \square^l} R_{\alpha \beta} R^{\alpha \beta} \right]=0\,,\label{EOM}
\end{eqnarray}
where $\square^{l,r}$ act on the left and right arguments (on the right of the incremental ratio) as indicated inside the brackets.

From (\ref{EOM}), one could find that the Ricci-flat solution to $R_{\mu\nu}=0$ is also an exact solution to $E_{\mu\nu}=0$~\cite{Li:2015bqa}.
It is   given by the Schwarzschild solution with line element
\begin{eqnarray}
 ds^2 = \bar{g}_{\mu\nu}d x^\mu d x^\nu =-f(r)dt^2 +\frac{1}{f(r)}dr^2+ r^2(d\theta^2 + \sin^2\theta d\phi^2),\label{sch}
\end{eqnarray}
where $f(r)$ is the metric function defined by
\begin{equation}
f(r)=1-\frac{r_0}{r}
\end{equation}
with the horizon radius (size)  $r_0$.
Furthermore, the Kerr metric, being another Ricci-flat  solution to $R_{\mu\nu}=0$, is also an  exact solution to  the non-local gravity.

\section{Perturbations: linearized equations}

Now, let us  derive the linearized equation from (\ref{EOM}) for the case of $\gamma_2\not=0$  by considering the perturbation $h_{\mu\nu}$ around the background metric tensor $\bar{g}_{\mu\nu}$,
\begin{equation}
g_{\mu\nu}=\bar{g}_{\mu\nu}+h_{\mu\nu},
\end{equation}
where overbar ( $\bar{}$ ) denotes the background spacetime.
First of all, we would like to mention  that the black hole solution  obtained from (\ref{action1}) with  $\gamma_2 =0$ by choosing either (\ref{ff1}) or (\ref{ff2}) is stable against linear perturbations~\cite{Calcagni:2017sov}.
When choosing the case of (\ref{ff1}), its linearized equation  is reduced to
\begin{equation} \label{lein-eq}
\delta R_{\mu\nu}(h)=0
\end{equation}
 implying the stability of the Schwarzschild black hole in Einstein gravity~\cite{Regge:1957td,Zerilli:1970se,Vishveshwara:1970cc,Kwon:1986dw}.
Eq.(\ref{lein-eq}) is indeed  a second-order differential equation,  which is solvable for $h_{\mu\nu}$.

On the other hand, for the case of (\ref{ff2}), its linearized equations are composed of  the two forms
\begin{eqnarray}
&&\left(\bar{\square}-M^2\right) \delta R(h)= 0, \label{lineq1} \\
&&  \delta R_{\mu\nu}(h)-\frac{1}{6}\bar{g}_{\mu\nu}\delta R(h) - \frac{1}{3 M^2} \bar{ \nabla}_{\mu} \bar{\nabla}_\nu \delta R(h) =0,\label{lineq2}
\end{eqnarray}
which are surely independent of the exponential form of entire function $e^{H(\bar{\square})}$.
Here, $\bar{\square}$ is the background d'Alembertian  defined by
\begin{eqnarray} \label{lapalcian}
\bar{\square}=\bar{g}^{\mu\nu}\bar{\nabla}_\mu \bar{\nabla}_\nu &=&-\frac{1}{f(r)}\frac{\partial^2}{\partial t^2}+\frac{1}{r^2}\frac{\partial}{\partial r}\Big(r^2f(r) \frac{\partial}{\partial r}\Big) \nonumber \\
&+&\frac{1}{ r^2\sin \theta}\frac{\partial}{\partial \theta}\Big(\sin \theta\frac{\partial}{\partial \theta}\Big)+\frac{1}{ r^2 \sin^2\theta}\frac{\partial^2}{\partial \phi^2}. \label{lapla1}
\end{eqnarray}
Thus, the linearized equations (\ref{lineq1}) and (\ref{lineq2}) are exactly the same one obtained from the local Starobinsky theory ${\cal L}_f=\sqrt{|g|}[ R+R^2/(6M^2)]$ in Ref.~\cite{Starobinsky:1980te}.
Eq.(\ref{lineq2}) corresponds to a second-order equation for $\delta R(h)$ coupled to $\delta R_{\mu\nu}(h)$, which is not easy to be solved.
In other words, Eq.(\ref{lineq2}) is a fourth-order equation for the metric perturbation $h_{\mu\nu}$.
Therefore,  the stability can be proved  by introducing an auxiliary field at the level of the action (by lowering ${\cal L}_f$ to a second order scalar-tensor theory) before performing perturbation process~\cite{Myung:2011ih}.

For case of $\gamma_2\not=0$,  one may attempt to  derive a more simpler equation of motion because Eq.(\ref{EOM}) is too lengthy to analyze the stability of the black hole.
Ignoring quadratic order in the Ricci tensor ($ {\bf Ric}$), one finds~\cite{Li:2015bqa}
\begin{equation}
G_{\mu\nu}+2\frac{\delta R_{\alpha\beta}(g)}{\delta g^{\mu\nu}}\Big[g^{\alpha\beta} \gamma_0(\square) R + \gamma_2(\square) R^{\alpha\beta}\Big]+ O({\bf Ric}^2) =0\,.\label{EOM1}
\end{equation}
We note  that all the complicated incremental ratios in Eq.(\ref{EOM}) are dropped out of Eq.(\ref{EOM1})
since these ratios are quadratic in the Ricci tensor.
 Considering the  linear perturbation  of $\delta R_{\mu\nu}( h)$, the replacement of  $\bar{R}_{\mu\nu} =0$ cancels them out.

Imposing the unitarity condition of $\gamma_2(\square)=-2\gamma_0(\square)$ around the Minkowski spacetime, one reduces  Eq.(\ref{EOM1})  to
\begin{equation}
G_{\mu\nu}+2\frac{\delta R_{\alpha\beta}(g)}{\delta g^{\mu\nu}}\gamma_2(\square) G^{\alpha\beta}+ O({\bf Ric}^2) =0\,.\label{EOM2}
\end{equation}
Using the relation
\begin{equation}
\frac{\delta R_{\alpha\beta}(g)}{\delta g^{\mu\nu}}=\frac{1}{2}g_{\alpha(\mu}g_{\nu)\beta}\square +\frac{1}{2}g_{\mu\nu} \nabla_\alpha \nabla_\beta -g_{\alpha (\mu|}\nabla_\beta \nabla_{|\nu)},
\end{equation}
we obtain an  equation of motion  as
\begin{eqnarray}
G_{\mu\nu}&+& \square\Big(\gamma_2(\square) G_{\mu\nu}\Big)+ \nabla_\alpha \nabla_\beta \Big(\gamma_2(\square) G^{\alpha\beta}\Big)g_{\mu\nu} \nonumber \\
&-&2g_{\alpha (\mu|}\nabla_\beta \nabla_{|\nu)}\Big(\gamma_2(\square) G^{\alpha\beta}\Big)+O({\bf Ric}^2) =0\,.\label{EOM3}
\end{eqnarray}
In the case of $\gamma_2(\square)=-1/m^2$, using the Bianchi identity of $ \nabla^{\mu}G_{\mu\nu}=0$ and
the commutation of covariant derivatives acting on the tensor~\cite{Li:2015bqa}
\begin{equation}\label{EOM-1}
\nabla_\beta\nabla_\nu G^{\beta}_\mu-\nabla_\nu\nabla_\beta G^{\beta}_\mu=R^\beta_\nu G_{\beta\mu}-R_{\alpha\mu\beta\nu}G^{\alpha\beta},
\end{equation}
one finds the fourth-order equation which contains a single Riemann tensor-term as
\begin{equation}\label{EOM-2}
\square G_{\mu\nu}+ 2 R_{\mu\alpha\nu\beta}G^{\alpha\beta}-2R_{(\mu}~^\beta G_{\nu)\beta}-m^2 G_{\mu\nu}=0.
\end{equation}
Here, we note  the appearance of the second  term in Eq.(\ref{EOM-2}) which is a necessary term to analyze the stability of the black hole in forth-order gravity.
However, for $\gamma_2(\square) =-2\gamma_0(\square)$, it is  important to note that
 Eq.(\ref{EOM3}) gets infinitely many Riemann tensor-terms  which are non-zero on the black hole background
 because the operator ($\nabla_\mu$) does not commute with the d'Alembertian ($\square$).
As a simple  example of $\gamma_2(\square)=\square$ (Lee-Wick gravity), one has to use the relation
\begin{eqnarray}
\nabla_\beta (\square G^{\alpha\beta})=\square (\nabla_\beta G^{\alpha\beta})
&-&\nabla^\gamma(R_{\beta\gamma}G^{\alpha \beta}) \nonumber \\
&-&\nabla^\gamma(R^{\alpha}~_{\tau\beta\gamma})G^{\tau \beta}+2R^{\alpha}~_{\tau\beta\gamma}\nabla^\gamma G^{\tau\beta}, \label{EOM-3}
\end{eqnarray}
where the first term of right-handed side in (\ref{EOM-3}) is zero due to the Bianchi identity, but two Riemann tensor-terms appear as the last two terms.
Hence, Eq.(\ref{EOM3}) is not  suitable for analyzing the stability of the black hole in the non-local gravity.

On the other hand, we  consider a  sixth-order gravity including the Lee-Wick gravity in~\cite{Accioly:2016qeb}.
For $\alpha=-\beta/2$, $A=-B/2$, and $2/\kappa^2=1$, a full equation takes the form
\begin{equation} \label{EOM-5}
 G_{\mu\nu}+\square (\alpha +A \square) G_{\mu\nu} +2(\alpha+A \square)  (R_{\mu\sigma\nu\rho} R^{\sigma\rho})+ {\cal O} ({\bf Ric}^2)=0,
\end{equation}
which involves a single Riemann tensor-term. In the case of $A=0$ and $\alpha=-1/m^2$, Eq.(\ref{EOM-5}) recovers Eq.(\ref{EOM-2}) because a part of the second term and the third term in Eq.(\ref{EOM-2}) belongs to ${\cal O} ({\bf Ric}^2)$.
Also, one finds the equation for the Lee-Wick gravity for $\alpha=0$.

Observing Eq.(\ref{EOM-5}), we  might deduce a full equation for the form factor $\gamma_2(\square)$ as
\begin{equation} \label{EOM-6}
 G_{\mu\nu}+\square (\gamma_2(\square) G_{\mu\nu}) +2\gamma_2(\square)  (R_{\mu\sigma\nu\rho} R^{\sigma\rho})+ {\cal O} ({\bf Ric}^2)=0.
 \end{equation}
However, we must caution the reader  that what we are deriving Eq.(\ref{EOM-6}) is only guesswork to illustrate the difficulties encountered in proving stability of AdS black hole in the non-local gravity.
Here, we note that  any other Riemann tensors are not generated except the third  term in (\ref{EOM-6}), compared to the infinitely many terms from Eq.(\ref{EOM3}).
Therefore, we may  use Eq.(\ref{EOM-6}) to obtain  the linearized equation instead of Eq.(\ref{EOM3}).
At this stage, we wish to comment that although the  third term disappears in the background equation because of $\bar{R}_{\mu\nu}=0$, it still survives in the linearized equation because of $\bar{R}_{\mu\alpha\nu\beta}\not=0$ and $\delta R^{\alpha\beta}(h)\not=0$.
Even for $\gamma_2=-2\gamma_0=-1/m^2$, this term plays an important role in performing the stability analysis of black hole in  fourth-order gravity~\cite{Whitt:1985ki,Myung:2013doa,Stelle:2017bdu,Lu:2017kzi}.

Taking into account $\bar{G}_{\mu\nu}=0$, let us  linearize   Eq.(\ref{EOM-6}) to be
\begin{equation}
\delta G_{\mu\nu}(h)+ \bar{\square}\Big[\gamma_2(\bar{\square}) \delta G_{\mu\nu}(h)\Big]+ 2\gamma_2(\bar{\square}) \Big[\bar{R}_{\mu\alpha\nu\beta}\delta R^{\alpha\beta}(h)\Big]=0\,\label{EOM4}
\end{equation}
with
\begin{equation}
\delta G_{\mu\nu}(h)=\delta R_{\mu\nu}(h)-\frac{1}{2}\bar{g}_{\mu\nu}\delta R(h).
\end{equation}
Furthermore, taking the trace of (\ref{EOM4}), we obtain  its trace equation
\begin{equation} \label{EOM5}
\bar{\square}\Big[\gamma_2(\bar{\square})\delta R(h)\Big]+\delta R=0.
\end{equation}
Consequently, we note that Eq.(\ref{EOM4}) is a linearized equation derived newly from Eq.(\ref{EOM-6}) for $\gamma_2=-2\gamma_0$.
Moreover,   Eq.(\ref{EOM4}) is quite different  from the linearized equation for $\gamma_2=0$ as~\cite{Calcagni:2017sov}
\begin{equation}
\delta G_{\mu\nu}(h)+2\Big(\bar{g}_{\mu\nu}\bar{\square}-\bar{\nabla}_{(\mu} \bar{\nabla}_{\nu)}\Big)\gamma_0(\bar{\square})\delta R(h)=0,
\end{equation}
which was used to derive Eqs.(\ref{lineq1}) and (\ref{lineq2}).

\section{Stability analysis}

In this section, we wish to investigate the stability of the Schwarzschild solution (\ref{sch}) in non-local gravity.
However, it seems difficult  to solve (\ref{EOM4}) directly.

\subsection{Prototype analysis: Lichnerowicz approach}

As a prototype analysis, we first consider  the case of $\gamma_2=-2\gamma_0=-1/m^2<0$, which corresponds to fourth-order  gravity~\cite{Whitt:1985ki,Myung:2013doa,Stelle:2017bdu,Lu:2017kzi}.
This corresponds to a non-unitary and renormalizable local gravity and hence, this gravity suffers from ghost problem~\cite{Stelle:1976gc}.
In the case of  $\gamma_2=-3\gamma_0=-1/m^2$, one obtains the Einstein-Weyl gravity which is a non-unitary and non-renormalizable local gravity.

In the non-unitary and renormalizable local theory, Eqs.(\ref{EOM4}) and (\ref{EOM5}) reduce to
\begin{equation}\label{EOM6}
\bar{\square} \delta G_{\mu\nu}(h)+ 2 \bar{R}_{\mu\alpha\nu\beta}\delta R^{\alpha\beta}(h)-m^2\delta G_{\mu\nu}(h)=0
\end{equation}
and
\begin{equation} \label{EOM7}
\bar{\square}\delta R-m^2\delta R=0,
\end{equation}
respectively.
Then, let us  introduce the traceless linearized Ricci tensor ($\delta R_{~~~\mu}^{{\rm T}\mu}(h)=0$)  as
\begin{equation}
  \delta R_{\mu\nu}(h)= \delta R_{\mu\nu}^{\rm T}(h)+\frac{\bar{g}_{\mu\nu}}{4}\delta R(h),~~  \delta G_{\mu\nu}(h)=\delta R_{\mu\nu}^{\rm T}(h)-\frac{\bar{g}_{\mu\nu}}{4}\delta R(h).
  \end{equation}
The linearized Einstein equation (\ref{EOM6}) together with  the  linearized trace equation (\ref{EOM7}) gives us a decoupled equation for $\delta R^{\rm T}_{\mu\nu}(h)$
\begin{equation}\label{EOM8}
\bar{\square} \delta R^{\rm T}_{\mu\nu}(h)+ 2 \bar{R}_{\mu\alpha\nu\beta}\delta R^{{\rm T} \alpha\beta}(h)-m^2\delta R^{\rm T}_{\mu\nu}(h)=0.
\end{equation}
For simplicity, we may use a Lichnerowicz-type argument~\cite{Lu:2017kzi},  which states that we multiply (\ref{EOM7}) by $\delta R(h)$ and integrate over the spatial 3-volume between the horizon and infinity.
With appropriate boundary conditions and for the positive mass-squared $m^2$, one may deduce the non-propagation of the linearized Ricci scalar ($\delta R(h)=0$).
However, it is not a correct statement to yield the non-propagation of the linearized Ricci scalar.
When considering the Einstein-Weyl gravity~\cite{Myung:2013doa}, it is very natural to obtain  $\delta R(h)=0$.

To make a further progress on the stability analysis, we wish to choose the non-propagation of the Ricci scalar here.
Using the linearized Bianchi identity
$\bar{\nabla}^\mu \delta G_{\mu\nu}(h)=0$ with $\delta R(h)=0$,
we obtain the transversality condition of  $ \bar{\nabla}^\mu \delta R_{\mu\nu}(h)=0$.  So, $\delta R_{\mu\nu}(h)$ might be  considered as  a transverse and traceless (TT) tensor.
In this case, $\delta R^{\rm TT}_{\mu\nu}$ describes five degrees of freedom (DOF) because it is a TT tensor in four dimensions.
Fortunately, the linearized equation (\ref{EOM8}) can be expressed as the Lichnerowicz eigenvalue equation as
\begin{equation}\label{EOM9}
\Big(\Delta_{\rm L}+m^2 \Big) \delta R^{\rm TT}_{\mu\nu}=0,~~~m^2=-\frac{1}{\gamma_2},
\end{equation}
where the Lichnerowicz operator on the Ricci-flat background is given by~\cite{Gibbons:2002pq}
\begin{equation} \label{lichnero}
\Delta_{\rm L} X_{\mu\nu}=-\bar{\square}X_{\mu\nu}-\bar{R}_{\mu\rho\nu\sigma}X^{\rho\sigma}.
\end{equation}
We note that Eq.(\ref{EOM9}) is a second-order differential equation for $\delta R^{\rm TT}_{\mu\nu}$.
It is known in Ref.~\cite{Myung:2013doa} that the time-dependent instability of the Schwarzschild black hole in fourth-order  gravity
is directly related to a  familiar instability of the five-dimensional black string as found by Gregory and Laflamme (GL)~\cite{Gregory:1993vy}.
The $s(\ell=0)$-mode analysis is suitable for investigating the massive
graviton propagation in the fourth-order gravity, but not for studying the
massless graviton propagation  in Einstein gravity. In general,
the $s$-mode analysis of the massive graviton  shows the
GL  instability~\cite{Gregory:1993vy}, which never
appears in either Einstein gravity or $f(R)$ gravity~\cite{Babichev:2013una,Brito:2013wya}.
In this case, the even-parity metric perturbation is used to define  a $s$-mode analysis in
the fourth-order gravity and whose form is given by $\delta
R^{\rm TT}_{tt},~\delta R^{\rm TT}_{tr},~\delta R^{\rm TT}_{rr}$ and $\delta R^{\rm TT}_{\theta\theta}$ as~\cite{Myung:2013doa}
\begin{eqnarray}
\delta R^{\rm TT}_{\mu\nu}(t,r,\theta)=e^{\Omega t} \left(
\begin{array}{cccc}
\delta R^{\rm TT}_{tt}(r) & \delta R^{\rm TT}_{tr}(r) & 0 & 0 \cr \delta
R^{\rm TT}_{tr}(r) & \delta R^{\rm TT}_{rr}(r) & 0 & 0 \cr 0 & 0 &  \delta
R^{\rm TT}_{\theta\theta}(r) & 0 \cr 0 & 0 & 0 & \sin^2\theta\delta
R^{\rm TT}_{\theta\theta}(r)
\end{array}
\right). \label{evenp}
\end{eqnarray}
Here we note that  $e^{\Omega t}$ represents a real exponential form, for which $\Omega>0$  corresponds to instability of  exponentially growing mode in time.
Eliminating all except $\delta R^{\rm TT}_{tr}$, Eq.(\ref{EOM9})
reduces to a second-order radial equation for $\delta R^{\rm TT}_{tr}$
\begin{equation} \label{secondG-eq} A
\frac{d^2}{dr^2}\delta R_{tr}^{\rm TT} +B\frac{d}{dr} \delta R_{tr}^{\rm TT}+C\delta
R^{\rm TT}_{tr}=0,
\end{equation}
where $A,B,$ and $C$ are given by Eqs.(36)-(38) in Ref.~\cite{Myung:2014nua}.
 It is worth noting
that the $s$-mode perturbation is described by single DOF ($\delta R^{\rm TT}_{tr}$)  but not five
DOF. The boundary conditions are chosen such  that $\delta R^{\rm TT}_{tr}$ should
be regular on the future horizon and vanishing at infinity.

Now, we are in a position to solve (\ref{secondG-eq}) numerically, and
find unstable modes. See Fig. 1,  which is  generated from the
numerical analysis. From the observation of Fig. 1 with ${\cal
O}(1)\simeq 0.86$, we find unstable modes~\cite{Babichev:2013una}
for a low-mass Schwarzschild black hole,  which satisfies
\begin{equation} \label{unst-con}
0<m<\frac{{\cal O}(1)}{r_0}. \end{equation}
\begin{figure*}[t!]
   \centering
   \includegraphics{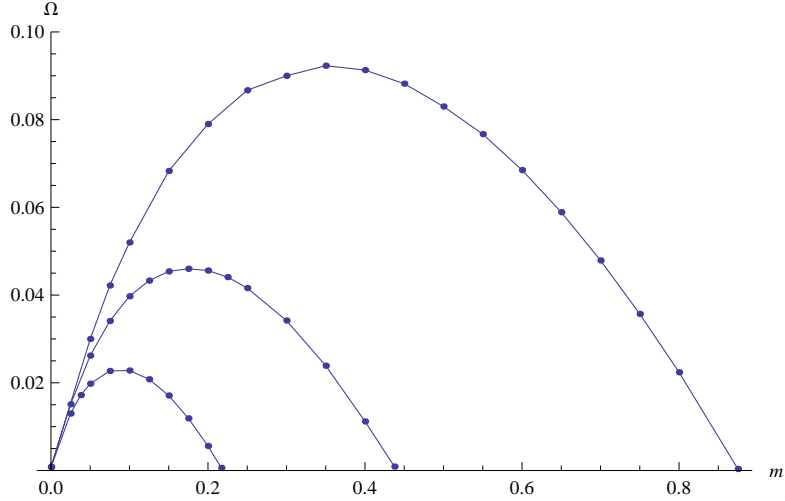}
\caption{Plots of unstable modes ($\bullet$) on three curves with the horizon radii $r_0=1,2,4$.
The $y(x)$-axis denote $\Omega(m)$, respectively. The smallest curve represents
$r_0=4$, the medium denotes $r_0=2$, and the largest one shows
$r_0=1$.}
\end{figure*}

As a consequence, it indicates  that the region of instability becomes
progressively smaller, as the horizon size $r_0$ increases.
This shows that the stability analysis of black hole obtained in fourth-order gravity with $\gamma_2=-2\gamma_0=-1/m^2$
being a non-unitary and renormalizable gravity is quite  different from that with $\gamma_2=0$ which is reduced  to Einstein or  $f(R)$ gravity.

Finally, we would like to mention that the Einstein-Weyl gravity  being a non-unitary and non-renormalizable theory  yields
the non-propagation of the linearized Ricci scalar  and thus, the stability analysis can be performed without any handicaps.

\subsection{Stability of black hole in the non-local gravity}

For the non-local gravity with $\gamma_2(\bar{\square})=-2\gamma_0(\bar{\square})$ in Eq.(\ref{ff1}),
plugging $\gamma_2=(e^{H(\bar{\square})}-1)/\bar{\square}$ into Eq.(\ref{EOM4}) leads to
\begin{equation} \label{EOM11}
e^{H(\bar{\square})}\delta R_{\mu\nu}(h)+ \frac{2(e^{H(\bar{\square})}-1)}{\bar{\square}} \Big( \bar{R}_{\mu\alpha\nu\beta}\delta R^{\alpha\beta}\Big)=0,
\end{equation}
which depends on the exponential form of entire function $e^{H(\bar{\square})}$. However, it is difficult to solve (\ref{EOM11}) directly.

For a simple  case of Lee-Wick gravity with $\gamma_2(\bar{\square})=\bar{\gamma}^2_2\bar{\square}$ (unitary and renormalizable local gravity~\cite{Modesto:2015ozb,Modesto:2016ofr})~\cite{Bambi:2016wmo}, its linearized equation around the Schwarzschild spacetime is given by
\begin{equation}\label{EOM12}
\bar{\square}\Big[\bar{\square} \delta R_{\mu\nu}(h)\Big]+ 2\bar{\square}\Big[\bar{R}_{\mu\alpha\nu\beta}\delta R^{ \alpha\beta}(h)\Big]+\frac{1}{\bar{\gamma}^2_2}\delta R_{\mu\nu}(h)=0,
\end{equation}
which is reduced to
\begin{equation}\label{EOM12-1}
\Delta_{\rm L} \delta R_{\mu\nu}(h)-\frac{1}{\bar{\gamma}^2_2\bar{\square}}\delta R_{\mu\nu}(h)=0.
\end{equation}
Unfortunately, this equation cannot be solved for stability analysis because it does not  represent the Lichnerowicz eigenvalue equation.
This implies that the fourth-order gravity may provide a best equation  to study the stability of  black hole when taking  the linearized  Ricci tensor as a physical tensor.
However, all higher-derivative gravities  more than fourth-derivative including (\ref{action1}) may be not suitable for  analyzing the black hole stability when employing the Lichnerowicz operator  approach.

Finally, one may consider  the absence of the Einstein-Hilbert term $R$ in the Lee-Wick gravity.
In this case, its action takes the form
\begin{equation} \label{action1v}
\tilde{S}_{\rm LW} = \frac{2}{\kappa^2}\int  d^4 x \sqrt{|g|} \left[G_{\mu\nu} \square R^{\mu\nu}\right].
 \end{equation}
The corresponding linearized equation  leads to
\begin{equation} \label{EOM13}
\bar{\square}\Big[\Delta_{\rm L} \delta R_{\mu\nu}(h)\Big]=0
\end{equation}
and   its trace equation is given by
\begin{equation} \label{EOM14}
 \bar{\square}^2\delta R(h)=0.
\end{equation}
In this case, the linearized equation (\ref{EOM13})  could lead to a purely second-order equation for $\delta R_{\mu\nu}$ (fourth-order  equation for $h_{\mu\nu}$) as
\begin{equation} \label{EOM15}
\Delta_{\rm L}\delta R_{\mu\nu}(h)=0.
\end{equation}
Then, one could obtain
\begin{equation}
\Delta_{\rm L}^2h_{\mu\nu}=0
\end{equation}
because  $\delta R_{\mu\nu}=\Delta_{\rm L}h_{\mu\nu}/2$ under the transverse and trace gauge ($\bar{\nabla}^\mu h_{\mu\nu}=0,~h=0$).

In 1970, Zerilli~\cite{Zerilli:1970se} proved that $\delta R(h)_{\mu\nu}=\Delta_{\rm L}h_{\mu\nu}/2=0$ could be casted into a Schr\"{o}dinger equation for single field (${\bf h}$) in Einstein gravity.
Then, it turns out that the Schwarzschild black hole is stable against the even-parity perturbations in Einstein gravity.
Similarly, it is straightforward to show   that Eq.(\ref{EOM15}) can be casted into a Schr\"{o}dinger equation for single field (${\cal R}$) in the Lee-Wick gravity without $R$ (\ref{action1v}) when replacing $h_{\mu\nu}$ by
$\delta R_{\mu\nu}$.  It is argued  that the black hole is stable against the Ricci-tensor perturbations in the Lee-Wick gravity without $R$.

\section{Discussions}

In this work, we have  investigated the linear stability of Schwarzschild black hole in the unitary and renormalizable  non-local gravity with $\gamma_2(\bar{\square})=-2\gamma_0(\bar{\square})$.
As a result, we have derived a linearized equation (\ref{EOM11}). However,  it is  a formidable task to solve (\ref{EOM11}) directly because of the exponential form of entire function.
So, we could not check that the Schwarzschild black hole exists really in  the unitary and renormalizable non-local gravity with  $\gamma_2(\bar{\square})=-2\gamma_0(\bar{\square})$.
Note that  if the black hole passes the stability test, one  saves that black hole.

One may have an early expectation that non-locality could resolve all the singularities of Einstein gravity and fourth-order gravity.
As a counter example,  it is known that a low-mass black hole is unstable against the Ricci tensor perturbations in fourth-order gravity.
If one considers the black hole found from the sixth-derivative gravity (Lee-Wick gravity: unitary and renormalizable local gravity), one  conjectures that this black hole may be stable.
However,  its linearized equation is given by
Eq.(\ref{EOM12-1}) which is not a Lichnerowicz eigenvalue equation.
The reason why we adhere to the Lichnerowicz operator approach here  is that there is no other way of analyzing the stability of black hole in higher-derivative gravity~\cite{Stelle:2017bdu}.
Actually, the inverse of d'Alembertian $\bar{\square}$ in the last term  of Eq.(\ref{EOM12-1}) makes the stability analysis less tractable.

Consequently, considering the simultaneous requirement for  unitarity and renormalizability  around the Minkowski background which implies the non-local gravity with $\gamma_2(\bar{\square})=-2\gamma_0(\bar{\square})$,
the exponential form of the entire function  makes the stability analysis of black hole difficult.
This means  that non-locality may not be a good tool to save the black hole solutions.  For this purpose, we mention that  the non-locality is insufficient to guarantee singularity  freedom,
but the conformal invariance seems to play a more important role~\cite{Calcagni:2010ab}.
Even though one needs more and more the d'Alembertian to have unitary and renormalzable gravity around the Minkowski spacetime,
increasing  these operators makes the stability analysis of black hole more obscured.

In summary, we have generalized the previous work~\cite{Calcagni:2017sov} about the stability of black hole soutions in non-local gravity.
As a result, we concluded  that what makes the problem difficult in stability analysis of black hole is the simultaneous requirement of  unitarity and renormalizability around the Minkowski spacetime.

\section*{Acknowledgement}
This work was supported by the National Research Foundation of Korea (NRF) grant funded by the Korea government (MOE)
 (No. NRF-2017R1A2B4002057).

\end{document}